# AN ANALYTIC MODEL FOR THE SPATIAL CLUSTERING OF DARK MATTER HALOES

H. J. Mo and S. D. M. White

Max-Planck-Institut für Astrophysik
Karl-Schwarzschild-Strasse 1
85748 Garching, Germany

Institute for Theoretical Physics
University of California
Santa Barbara, CA 93106-4030, USA

## ABSTRACT

We develop a simple analytic model for the gravitational clustering of dark matter haloes to understand how their spatial distribution is biased relative to that of the mass. The statistical distribution of dark haloes within the initial density field (assumed Gaussian) is determined by an extension of the Press-Schechter formalism. Modifications of this distribution caused by gravitationally induced motions are treated using a spherical collapse approximation. We test this model against results from a variety of N-body simulations, and find that it gives an accurate description of a bias function, $b(M, R, \delta) = \delta_{\rm h}(M, R, \delta)/\delta$, where $\delta_{\rm h}(M, R, \delta)$ is the mean overdensity of haloes of mass $M$ within spheres which have radius $R$ and *mass* overdensity $\delta$; the results depend only very weakly on how haloes are identified in the simulations. This bias function is sufficient to calculate the cross-correlation between dark haloes and mass, and again we find excellent agreement between simulation results and analytic predictions. Because haloes are spatially exclusive, the variance in the count of objects within spheres of fixed radius and overdensity is significantly smaller than the Poisson value. This seriously complicates any analytic calculation of the autocorrelation function of dark halos. Our simulation results show, however, that this autocorrelation function is proportional to that of the mass over a wide range in $R$, even including scales where both functions are significantly greater than unity. Furthermore, the constant of proportionality is very close to that predicted on large scales by the analytic model. Since analytic formulae for the nonlinear autocorrelation function of the mass are already known, this result permits an entirely analytic estimate of the autocorrelation function of dark haloes. We use our model to study how the distribution of galaxies may be biased with respect to that of the mass. In conjunction with other data these techniques should make it possible to measure the amplitude of cosmic mass fluctuations and the density of the Universe.

**Key words:** galaxies: clustering-galaxies: formation-cosmology: theory-dark matter

## 1 INTRODUCTION

A fundamental problem in cosmology is to understand how the spatial distribution of galaxies (and of galaxy clusters) is related to that of the underlying mass. In the standard scenario of structure formation, a dominant dissipationless component of dark matter is assumed to aggregate into dark matter clumps, the virialized parts of which are usually called dark haloes. Galaxies then form by the cooling and condensation of gas within these dark haloes (White and Rees 1978). Complex nongravitational processes which cannot be modelled reliably are likely to be critical in determining the properties of individual galaxies, yet they have little effect on the formation and clustering of dark haloes. As a result it is useful to approach the problem of galaxy biasing by first understanding how dark haloes are distributed relative to the mass.

Since dark haloes are highly nonlinear objects, their formation and evolution has traditionally been studied using N-body simulations (e.g. Frenk 1991; Gelb & Bertschinger 1994a,b and references therein). Such simulations are limited both in resolution and in dynamical range and can be difficult to interpret. Our understanding of their results could be substantially enhanced by simple physical models and the analytic approximations they provide. In particular, a simple and accurate analytic model could be used not only to carry out large parameter studies, but also to help reconstruct the mass distribution from observations. The present paper attempts to provide such a model.

The initial distribution of density fluctuations in the universe is usually assumed to be Gaussian, and so to be described completely by its power spectrum. This, in turn, is derived from a model for the origin of structure in the early universe (e.g. Kolb & Turner 1990). It is perhaps feasible to associate dark haloes with certain specific regions of the initial density field and to consider how these regions cluster as a result of the statistics of the initial conditions and of the motions induced by gravity. Kaiser (1984) used this idea to show how the strong clustering of Abell clusters could be explained using the statistics of high peaks in an initial Gaussian field. His formalism was developed extensively by Bardeen et al. (1986). These authors showed that if galaxies can be associated with high peaks of the initial density field then they should be more clustered than the mass, an effect usually called "galaxy biasing". Unfortunately it is not known how well galaxies correspond to high peaks of the initial field, and there is direct evidence that the correspondance of peaks with dark haloes is not particularly good (Frenk et al. 1988; Katz, Quinn & Gelb 1993). In particular, it is unclear how to deal with the problems that a single dark halo often contains several peaks, and that the present-day clustering of peaks differs substantially from that in the initial (Lagrangian) space as a result of gravitationally induced motions. Substantial progress in overcoming these difficulties has recently been made by Bond and Myers (1995a,b) in their "Peak-Patch Picture". In the present paper, however, we follow a less rigorous but simpler and more easily implemented route.

Press & Schechter (1974, hereafter PS) developed a formalism which identifies haloes at any given cosmic time with regions of the initial density field which just collapse at that time according to a spherical infall model. This theory can be extended so that it predicts not only the evolution of the mass function of dark haloes, but also the full statistical properties of the heirarchical clustering process (Bond et al. 1991; Bower 1991; Lacey & Cole 1993; Kauffmann & White 1993). Comparisons with N-body simulation data show detailed agreement for a very wide range of statistical properties of the clustering process (e.g. Bond et al. 1991; Bower 1991; Kauffmann & White 1993; and particularly, Lacey & Cole 1994). This agreement is quite surprising since the basic hypothesis underlying the PS approach is found to work poorly on an object by object basis (Bond et al. 1991; White 1995).

The PS theory developed in the above papers does not provide a model for the spatial clustering of dark haloes, but it is easily extended to construct such a model. We use the standard PS formalism both to define dark haloes from the initial density field, and to specify how their mean abundance within a large spherical region is modulated by the linear mass overdensity in that region. We then treat the gravitationally induced evolution of clustering by assuming that each region evolves as if spherically symmetric. Section 2 lays out this model in detail and shows how it can be used to calculate statistical properties of the clustering of dark haloes as a function of their mass and of the epoch at which they are identified. Section 3 then presents detailed tests of these predictions against a variety of large N-body simulations. Finally, in section 4 we discuss how our model might be used to understand biasing of the galaxy distribution with respect to that of the mass, and how these methods may help in reconstructing the cosmic mass distribution from observations.

## 2 THE MODEL

Although the model described here may readily be extended to other cosmologies, the present development assumes, for simplicity, an Einstein-de Sitter universe (i.e. that the total mass density parameter $\Omega = 1$, and the cosmological constant $\Lambda = 0$).

### 2.1 Initial density field and dark matter haloes

We assume that the initial overdensity field $\delta(\mathbf{x}) \equiv [\rho(\mathbf{x}) - \bar{\rho}]/\bar{\rho}$ (whose Fourier transform is denoted by $\delta_{\mathbf{k}}$) is Gaussian and is described by a power spectrum $P(k)$. The field $\delta(\mathbf{x})$ can be smoothed by convolving it with a spherical symmetric window function $W(r; R)$ having *comoving* characteristic radius $R$ (measured in current units). The smoothed field is

$$\delta(\mathbf{x}; R) = \int W(|\mathbf{x} - \mathbf{y}|; R)\delta(\mathbf{y})d^3y$$

$$= \int \hat{W}(k; R)\delta_{\mathbf{k}} exp(i\mathbf{k} \cdot \mathbf{x})d^3k, \qquad (1)$$

where $\hat{W}(k; R)$ is the Fourier transform of the window function $W(r; R)$. A useful quantity characterising the power spectrum is the rms fluctuation of mass in a given smoothing window:

$$\Delta^2(R) = \langle[\delta(\mathbf{x}; R)]^2\rangle = \int P(k)\hat{W}^2(k; R)d^3k. \qquad (2)$$

For a given window function the smoothed field $\delta(\mathbf{x}; R)$ is Gaussian and so has the following one-point distribution function

$$p(\delta; R)d\delta = \frac{1}{(2\pi)^{1/2}} \exp\left[-\frac{\delta^2}{2\Delta^2(R)}\right] \frac{d\delta}{\Delta(R)}. \tag{3}$$

Since both $\delta$ and $\Delta(R)$ grow with time in the same manner in linear perturbation theory, it is convenient to use their values linearly extrapolated to the present time. It is clear that these extrapolated quantities still obey equation (3). In what follows, we write our formulae in terms of the extrapolated quantities, unless otherwise stated. Also we will omit writing explicitly the smoothing radius $R$, but we will often use subscripts to distinguish $\Delta$, and other quantities, at different smoothing lengths [e.g. $\Delta_0 \equiv \Delta(R_0)$, $\Delta_1 \equiv \Delta(R_1)$]. For a top-hat window function, which we adopt throughout this paper, the average mass contained in a window of radius $R$ is simply $\bar{M}(R) = (4\pi/3)\bar{\rho}R^3$, where $\bar{\rho}$ is the mean density of the universe. For a given power spectrum $P(k)$, the quantities $R$, $\Delta$ and $\bar{M}$ are equivalent variables.

We will assume that dark halos are spherically symmetric, virialized clumps of dark matter. In an Einstein-de Sitter universe a spherical perturbation of linear overdensity $\delta$ collapses at redshift $z_c = \delta_c/\delta - 1$, where $\delta_c \equiv 1.686$. It is usually assumed that a collapsing structure virializes at half its radius of maximum expansion, implying a density contrast at the time of collapse of about 178. The mass $M_1$ of a halo is related to the initial comoving radius $R_1$ of the region from which it formed by

$$M_1 = \frac{4\pi}{3}\bar{\rho}R_1^3. \tag{4}$$

Note that we will always label the properties of dark haloes ($R$, $M$, $\Delta$ etc.) using the subscripts 1, 2, ... We will reserve the subscript 0 for the properties of uncollapsed spherical regions.

According to PS theory, the probability that a random mass element is part of a dark halo of mass exceeding $M_1$ at some given redshift $z_1$ is just twice the probability that a surrounding sphere of mass $M_1$ in the initial conditions has linearly extrapolated overdensity greater than $\delta_c$ at that redshift. This probability is

$$F(M_1, z_1) = \int_{(1+z_1)\delta_c}^{\infty} p(\delta; R_1)d\delta, \tag{5}$$

where $p(\delta; R_1)$ is given by equation (3). This equation can be rewritten as

$$F(M_1, z_1) = F(\nu_1) = erfc\left[\frac{\nu_1}{\sqrt{2}}\right], \tag{6}$$

where $\nu_1 \equiv \delta_1/\Delta_1$, we define $\delta_1 = (1+z_1)\delta_c$, and $erfc(x)$ is the complementary error function. The differential mass distribution is then

$$f(M_1, z_1)dM_1 = -\frac{\partial F}{\partial M_1}dM_1 = \frac{2}{(2\pi)^{1/2}}\frac{\delta_1}{\Delta_1^2}\exp\left[-\frac{\delta_1^2}{2\Delta_1^2}\right]\frac{d\Delta_1}{dM_1}dM_1. \quad (7)$$

Hence the comoving number density of haloes, expressed in current units, as a function of $M_1$ and $z_1$ is

$$n(M_1, z_1)dM_1 = -\left(\frac{2}{\pi}\right)^{1/2}\frac{\bar{\rho}}{M_1}\frac{\delta_1}{\Delta_1}\frac{d\ln\Delta_1}{d\ln M_1}\exp\left[-\frac{\delta_1^2}{2\Delta_1^2}\right]\frac{dM_1}{M_1}, \quad (8)$$

where $\bar{\rho}$ is the current mean density of the universe. Notice that in this theory a class of haloes must be defined by specifying *both* their mass $M_1$ (or equivalently $R_1$ or $\Delta_1$) and their redshift of identification $z_1$ (or equivalently $\delta_1$).

We now need formulae which relate halo abundances to the density field on larger scales. Bond et al. (1991) derive the probability that the overdensity at a randomly chosen point is $\delta_0$ when the initial density field is smoothed on scale $R_0$ *and* does not exceed $\delta_1$ for any larger smoothing scale:

$$q(\delta_0, \delta_1; R_0)d\delta_0 = \frac{1}{(2\pi)^{1/2}}\left[\exp\left(-\frac{\delta_0^2}{2\Delta_0^2}\right) - \exp\left(-\frac{(\delta_0 - 2\delta_1)^2}{2\Delta_0^2}\right)\right]\frac{d\delta_0}{\Delta_0}, \quad (9)$$

for $\delta_0 < \delta_1$ and $q = 0$ otherwise. In line with their reinterpretation of PS theory, they consider this to be the probability that a spherical region of initial radius $R_0$ has linear overdensity $\delta_0$ and is not contained in a collapsed object of mass exceeding $M_0$ at redshift $z_1$ given by $\delta_1 = (1+z_1)\delta_c$. Notice that our subscript convention means that $\delta_0$ refers to an uncollapsed region and so should be interpreted as the linear overdensity of that region extrapolated to the present, whereas $\delta_1$, $\delta_2$, etc. apply to collapsed halos and so should interpreted as $\delta_c$ times 1 plus the redshift at which each halo is identified.

Bond et al. (1991) also extend their PS argument to show that the fraction of the mass in a region of initial radius $R_0$ and linear overdensity $\delta_0$ which at redshift $z_1$ is contained in dark haloes of mass $M_1$ (where by definition $M_1 < M_0$) is given by

$$f(\Delta_1, \delta_1|\Delta_0, \delta_0)\frac{d\Delta_1^2}{dM_1}dM_1 = \frac{1}{(2\pi)^{1/2}}\frac{\delta_1 - \delta_0}{(\Delta_1^2 - \Delta_0^2)^{3/2}}\exp\left[-\frac{(\delta_1 - \delta_0)^2}{2(\Delta_1^2 - \Delta_0^2)}\right]\frac{d\Delta_1^2}{dM_1}dM_1. \quad (10)$$

Thus the average number of $M_1$ haloes identified at redshift $z_1$ in a spherical region with comoving radius $R_0$ and overdensity $\delta_0$ is

$$\mathcal{N}(1|0)dM_1 \equiv \frac{M_0}{M_1}f(1|0)\frac{d\Delta_1^2}{dM_1}dM_1, \quad (11)$$

where $f(1|0) \equiv f(\Delta_1, \delta_1 | \Delta_0, \delta_0)$. Notice that since $M_1$ is identified as a collapsed halo at $z_1 > 0$ whereas $M_0$ is assumed uncollapsed at z=0, we have $\delta_1 > \delta_0$. Equations (9) to (11) turn out to be sufficient for us to derive some interesting results for the pattern of halo clustering "imprinted" on the initial conditions as a result of the statistical properties of gaussian fields.

## 2.2 Clustering of haloes in Lagrangian space

From the preceding analysis it is clear that the number of halos of mass $M_1$, identified at redshift $z_1$, which form from the matter initially contained within spheres of radius $R_0$ and linear overdensity $\delta_0$ has a significant dependence on $\delta_0$. It is useful to quantify this by calculating the average overabundance of halos in such spheres relative to the global mean halo abundance. This is simply

$$\delta_h^L(1|0) = \frac{\mathcal{N}(1|0)}{n(M_1, z_1) V_0} - 1, \qquad (12)$$

where $V_0 = 4\pi R_0^3/3$, and the other quantities are taken from equations (8) and (11). This expression becomes particularly simple when the mass contained in the larger region is much greater than that of the haloes considered. When $R_0 \gg R_1$ (so that $\Delta_0 \ll \Delta_1$) and $|\delta_0| \ll \delta_1$, we have

$$\delta_h^L(1|0) = \frac{\nu_1^2 - 1}{\delta_1} \delta_0, \qquad (13)$$

where again $\nu_1 \equiv \delta_1/\Delta_1$. Thus the halo overdensity in these Lagrangian spheres is directly proportional to the linear mass overdensity. The constant of proportionality is the same as the one obtained from a related argument (sometimes called the peak-background split) by Efstathiou et al. (1988) and Cole & Kaiser (1989). It is useful to define a characteristic mass $M_*(z_1)$ for the nonlinear clustering at redshift $z_1$ through the requirement

$$\Delta(M_*) = \delta_1 = \delta_c(1 + z_1). \qquad (14)$$

Equation (13) then shows that haloes with mass $M_1$ exceeding $M_*$ are initially biased towards regions of positive linear overdensity, while those with $M_1 < M_*$ are initially biased to regions of negative linear overdensity. Notice also that while the positive bias factor can be very large for $M_1 \gg M_*$ the antibias cannot be more negative than $\delta_h^L/\delta_0 = -1/\delta_1 \geq -0.59$.

By combining equations (9) and (12) we can immediately calculate a measure of the cross-correlation between the number of dark halos and the amount of mass in Lagrangian spheres of radius $R_0$. We define such a measure by

$$\bar{\xi}_{hm}^L(R_0, M_1, z_1) = \langle \delta_h^L(1|0) \delta_0 \rangle_{R_0} = \frac{1}{n(M_1, z_1) V_0} \int_{-\infty}^{\infty} \delta_0 \, \mathcal{N}(1|0) q(\delta_0, \delta_1; R_0) d\delta_0, \qquad (15)$$

where $\langle \cdot \cdot \rangle_{R_0}$ denotes an average over all Lagrangian spheres of radius $R_0$ that are not contained in a collapsed object at redshift $z_1$. We use the notation $\bar{\xi}_{\rm hm}^{\rm L}$ because this definition gives an average of the standard cross-correlation between halos "h" and mass "m" in Lagrangian space "L". This average is carried out over a sphere of radius $R_0$ according to $\bar{\xi}(R_0) = V_0^{-2} \int\int \xi(|{\bf x} - {\bf y}|) d^3x d^3y$. We refer to $\bar{\xi}_{\rm hm}^{\rm L}$ and similar quantities as "average" (cross)-correlation functions. Notice that equation (15) requires only $M_1 < M_0$. It nowhere assumes that $\delta_0$ or $\delta_{\rm h}^{\rm L}$ should be small. We will show below that it provides a good description of our simulation data well into the nonlinear regime $\bar{\xi}_{\rm hm}^{\rm L} > 1$. On large scales, however, it simplifies using equation (13) to give

$$\bar{\xi}_{\rm hm}^{\rm L}(R_0, M_1, z_1) = \frac{(\nu_1^2 - 1)^2}{\delta_1^2} \Delta_0^2.$$

Notice that this quantity measures clustering in the pattern of dark halo formation sites at early times when the mass is almost uniform. We now turn to measures of clustering in the current universe where the positions of dark halos have been modified as a result of gravitationally induced motions.

### 2.3 Dynamical evolution of clustering

To model the clustering of dark halos at recent epochs we have to be able to calculate their expected abundance in spheres which at the desired redshift $z$ have radius $R$ and (possibly) nonlinear overdensity $\delta$. We relate these quantities to the initial Lagrangian radius $R_0$ and the extrapolated linear overdensity $\delta_0$ of the last section by using a spherical collapse model. In such a model each spherical shell moves as a unit and different shells do not cross until very shortly before they collapse through zero radius. Thus the mass interior to each shell is constant, giving $(1 + \delta)R^3 = R_0^3$. Furthermore, since dark halos in our PS model are defined to be objects identified at some specific redshift, the mean abundance of equation (11) can be taken as referring to halos of mass $M_1$ identified at redshift $z_1$ within spheres of radius $R(R_0, \delta_0, z_1)$ and overdensity $\delta(\delta_0, z_1)$.

For a spherical perturbation in an Einstein-de Sitter universe, the physical radius $R$ of a mass shell which had initial Lagrangian radius $R_0$ and mean linear overdensity $\delta_0$ is given for $\delta_0 > 0$ by

$$\frac{R(R_0, \delta_0, z)}{R_0} = \frac{3}{10} \frac{1 - \cos\theta}{|\delta_0|}; \tag{16}$$

$$\frac{1}{1+z} = \frac{3 \times 6^{2/3}}{20} \frac{(\theta - \sin\theta)^{2/3}}{|\delta_0|}. \tag{17}$$

For $\delta_0 < 0$, we just replace $(1 - \cos\theta)$ in equation (16) by $(\cosh\theta - 1)$ and $(\theta - \sin\theta)$ in equation (17) by $(\sinh\theta - \theta)$. Without loss of generality, let us assume $z = 0$ at the time when the clustering of haloes is examined. Then $\delta_0$ depends only on the present mass overdensity $\delta \equiv (R_0/R)^3 - 1$. The relation between these two quantities can be approximated accurately by

$$\delta_0 = -1.35(1+\delta)^{-2/3} + 0.78785(1+\delta)^{-0.58661} - 1.12431(1+\delta)^{-1/2} + 1.68647. \tag{18}$$

This interpolation formula has the correct asymptotic behaviour near $\delta = -1$ and as $\delta \to \infty$, as well as in the vicinity of $\delta = 0$.

Under the above assumption the average overdensity of dark haloes in spheres with current radius $R$ and current mass overdensity $\delta$ can be obtained immediately from equations (8) and (11):

$$\delta_{\rm h}(1|0) = \frac{\mathcal{N}(1|0)}{n(M_1, z_1)V} - 1, \tag{19}$$

where $V = 4\pi R^3/3$, $R_0 = R(1+\delta)^{1/3}$, and $\delta_0$ is determined from $\delta$ using equation (18). When $R_0 \gg R_1$ and $|\delta_0| \ll \delta_1$, we have

$$\delta_{\rm h}(1|0) = b(M_1, z_1)\delta = (1 + \frac{\nu_1^2 - 1}{\delta_1})\delta. \tag{20}$$

Again we find that halo overdensity is directly proportional to mass overdensity but the constant of proportionality (usually known as the linear bias parameter) is now always positive. The first term in the parenthesis defining $b(M_1, z_1)$ comes from the contraction (or expansion) of the spherical region, while the second reflects the bias in the initial density field as given by equation (13). It is interesting to note that for equation (20) to be valid, it is not necessary to have $\delta \ll 1$. Indeed, for haloes identified at redshifts of one or greater we show below that equation (20) can hold for $\delta$ substantially greater than unity. We also note that for $M_1 = M_*(z_1)$ we have $\nu_1 = 1$ and so $b = 1$; thus $M_*$ haloes are predicted to be unbiased relative to the mass.

In analogy with the analysis of the last section we can now define an average cross-correlation between dark haloes and mass, this time for spheres of fixed radius $R$ at $z = 0$,

$$\bar{\xi}_{\rm hm}(R, M_1, z_1) = \langle \delta_{\rm h}(1|0)\delta \rangle_R = \frac{1}{n(M_1, z_1)V} \int_{-\infty}^{\infty} \delta\, \mathcal{N}(1|0) p(\delta; R) d\delta, \tag{21}$$

where $\langle \cdot \cdot \rangle_R$ denotes an average over all spheres with radius $R$ at $z = 0$, $p(\delta; R)$ is the probability distribution function (PDF) of the mass overdensity in such spheres, and $R_0$ and $\delta_0$ in $\mathcal{N}(1|0)$ are related to $R$ and $\delta$ by the spherical collapse model. There have been many attempts to model the nonlinear PDF (often called the counts-in-cells distribution) for a given initial spectrum of gaussian density fluctuations (e.g. Bernardeau 1994, Colombi 1994). Unfortunately, the models proposed so far work reasonably well only in the linear and quasilinear regimes $\delta \lesssim 1$. In testing our model, we will often use a PDF derived directly from our N-body simulations. However, for illustration, we will also show results obtained using a simple lognormal approximation to the PDF (Coles & Jones 1991):

$$p(\delta; R)d\delta = \frac{1}{(2\pi)^{1/2}\sigma_l} \exp\left[-\frac{(\ln \rho + \sigma_l^2/2)^2}{2\sigma_l^2}\right] \frac{d\delta}{\rho}, \quad (22)$$

where $\rho = (1+\delta)$, $\sigma_l^2 = \ln[1+\sigma^2(R)]$, and $\sigma(R)$ is the *rms* overdensity fluctuation in a sphere of radius $R$. The latter can be obtained from the initial power spectrum through the formula given by Jain, Mo & White (1995). It turns out that such a PDF works remarkably well on scales where the average mass correlation function $\bar{\xi}_m(R) \lesssim 1$.

### 2.4. The autocorrelation functions of dark haloes

In analogy with the procedures of the last section we can define an average autocorrelation function at $z = 0$ for haloes of mass $M_1$ identified at redshift $z_1$ by

$$\bar{\xi}_{\rm hh}(R, M_1, z_1) = \langle [\delta_{\rm h}(1|0)]^2 \rangle_R = \frac{\int_{-\infty}^{\infty} [\mathcal{N}(1|0)]^2 p(\delta; R)d\delta}{[n(M_1, z_1)V]^2} - 1, \quad (23)$$

where, as before, $R_0$ and $\delta_0$ in $\mathcal{N}(1|0)$ are related to $R$ and $\delta$ by the spherical collapse model. In the limit where $R_1 \ll R_0$ and $|\delta_0| \ll \delta_1$ this gives

$$\bar{\xi}_{\rm hh}(R, M_1, z_1) = [b(M_1, z_1)]^2 \bar{\xi}_m(R), \quad (24)$$

where $\bar{\xi}_m(R)$ is the average mass correlation function. If $\delta_1$ is large (i.e. $z_1 > 1$), equation (24) follows from eq.(23), even when $\bar{\xi}_m(R) \gtrsim 1$.

It is important to note that the average correlation function $\bar{\xi}_{\rm hh}$ defined by equation (23) is not the same as the conventional one based on the variance of counts in randomly placed spheres (e.g. Peebles 1980, §36). This is because $\bar{\xi}_{\rm hh}(R)$ does not include the scatter of halo counts among spheres which have the same mean mass overdensity as well as the same radius. Let us denote the conventional average autocorrelation function by $\sigma_{\rm hh}^2$. Then

$$\sigma_{\rm hh}^2(R) = V^{-2} \int\int \xi_{\rm hh}(|\mathbf{x}-\mathbf{y}|)d^3x d^3y = \bar{\xi}_{\rm hh}(R) + \frac{\langle \mu(R,\delta) \rangle_R}{[n(M_1,z_1)V]^2} - \frac{1}{n(M_1,z_1)V}, \quad (25)$$

where $\mu(R,\delta)$ is the mean square scatter of halo counts in spherical regions with radius $R$ and overdensity $\delta$. The discreteness term in equation (25) (the last on its *rhs*) cancels the second term if the scatter among the counts in spheres of the same $R$ and $\delta$ (and so the same $R_0$ and $\delta_0$) is Poisson. However, since haloes are spatially exclusive this is not the case, and we expect that $\langle \mu(R,\delta)\rangle_R$ will be significantly less than $n(M_1,z_1)V$ for small $R$. Our $\bar{\xi}_{\rm hh}(R)$ is then not equivalent to $\sigma_{\rm hh}^2(R)$.

We will see below that halo exclusion effects are indeed important when count variances are estimated for massive haloes in N-body simulations. They modify the values of $\sigma_{\rm hh}^2(R)$ out to radii that are much larger than the typical sizes of haloes, giving a result which is systematically shallower than either $\bar{\xi}_{\rm hh}(R)$ or $\bar{\xi}_{\rm m}(R)$. Thus, without a proper understanding of the scatter $\mu(R,\delta)$, the usefulness of our model for predicting $\sigma_{\rm hh}^2$ is limited. This also means that $\sigma_{\rm hh}^2$ is related to the mass correlation function in a more complicated way than is suggested by equations (23) and (24). In principle, the scatter could be modelled by examining the formation histories of individual spherical regions of radius $R$ and overdensity $\delta$, as one does in studying the merging histories of individual dark haloes (e.g. Kauffmann and White 1993). This would, however, lead to a much more complicated model. In this paper, we will adopt a different approach.

Halo-halo exclusion must lead to a deficit of pairs at separations comparable to the sizes of haloes, but the pair count at larger separations could plausibly be almost unaffected. Thus the standard two-point correlation of haloes (as opposed to the *average* correlation functions we have been considering so far) may be insensitive to exclusion effects for large separations. To see this more clearly, recall that the two-point correlation function of haloes may be estimated as

$$\xi_{\rm hh}(R) = \frac{\Delta P(R)}{4\pi n R^2 \Delta R} - 1, \qquad (26)$$

where $\Delta P(R)$ is the average number of neighbours of a randomly chosen halo in the separation interval $R \pm \Delta R/2$ and $n$ is the mean number density of haloes. Now imagine splitting every halo into two halves each of which is counted separately. It is clear from equation (26) that $\xi_{\rm hh}(R)$ is unchanged for all $R$ larger than the splitting radius. The situation is quite different for $\sigma_{\rm hh}^2(R)$. From the definition of this quantity we can write (see Peebles 1980, §36)

$$\sigma_{\rm hh}^2(R) = \frac{\langle N^2(R)\rangle}{(nV)^2} - 1 - \frac{1}{nV}, \qquad (27)$$

where $\langle N^2(R)\rangle$ is the mean square count of haloes in spheres of radius $R$ and again $V = 4\pi R^3/3$. If we try splitting each halo in this case, the first two terms on the *rhs* of equation (27) are unchanged, while the last is reduced by a factor of two. Thus when $nV$ is small or the correlation signal is weak, $\sigma_{\rm hh}^2(R)$ can be affected significantly by small scale clustering. Based on this, it seems preferable to use $\xi_{\rm hh}(R)$ rather than $\sigma_{\rm hh}^2(R)$ to measure the autocorrelation of haloes on large scales; the effects of halo exclusion should then be reduced.

Unfortunately, as we have seen, our model based on the extended PS formalism does not lead directly to a model for $\xi_{\rm hh}$, or even for $\sigma_{\rm hh}^2(R)$. However, on large scales, we expect that the linear bias relation of equation (24) will apply also to $\xi_{\rm hh}(R)$. We can then write

$$\xi_{\rm hh}(R) = [b(M_1, z_1)]^2 \, \xi_{\rm m}(R), \qquad (28)$$

where $\xi_{\rm m}(R)$ is the standard autocorrelation function for the mass. When haloes with a range of masses are considered, $b(M_1, z_1)$ should be replaced in this equation by the value obtained by averaging it over $M_1$ with a weighting of $n(M_1, z_1)$. As we have discussed before, the linear bias relations of equations (20) and (24) are valid if $R_0 \gg R_1$ and $|\delta_0| \ll \delta_1$. Hence, equation (28) may hold even for $\xi_{\rm m} > 1$, especially when $z_1 \gtrsim 1$. In subsection 3.4, we will show that the autocorrelation of haloes in our simulations is described quite accurately by equation (28), even in the nonlinear regime of $\xi_{\rm m}$. This is an important result, because it means not only that $\xi_{\rm hh}(R)$ is very simply related to $\xi_{\rm m}$, but also that it can be estimated purely analytically using the analytic formulae for $\xi_{\rm m}$ given by Jain et al. (1995).

## 3 COMPARISON WITH NUMERICAL SIMULATIONS

We now test our analytic theory by detailed comparison with the results from a series of large cosmological N-body simulations. These simulations were performed using the particle-particle/particle-mesh (P$^3$M) code described by Efstathiou et al. (1988) and are very similar to the simulations of that paper. However, they are substantially larger ($N = 10^6$) and have higher resolution (gravitational softening length equal to $L/2500$, where $L$ is the side of the fundamental cube of the periodic simulation region). The initial conditions for each simulation imposed growing mode fluctuations corresponding to a Gaussian random field with power-law fluctuation spectrum, $P(k) \propto k^n$, onto a uniform "glass-like" initial particle load (see White 1995). All models assumed an Einstein-de Sitter universe and so their time evolution is expected to be self-similar. The initial power spectrum was normalized as described by Efstathiou et al. (1988) and "time" is measured by expansion factor $a$ since the start of the simulation ($a = 1$ for the initial conditions). Each model evolved further than those of Efstathiou et al. (until the largest virialized clusters contained more than $10^4$ partices) and a repetition of the tests of that paper showed that similarity scalings are accurately obeyed throughout the evolution. Some examples of such tests can be found in Jain et al. (1995) where correlation functions and power spectra for these same simulations were analysed.

For the purposes of this paper we need to define catalogues of dark haloes in the simulations. For most of our discussion we will use catalogues constructed by using the standard 'friends-of-friends' (FOF) group finder with a linkage length equal to 20% of the mean interparticle distance (e.g. Davis et al. 1985). This algorithm is easy to implement and has been extensively tested against the PS mass function (see Lacey & Cole 1994 for a careful discussion). For comparison, however, we will also present some results obtained using the spherical-overdensity (SO) grouping algorithm invented by Lacey & Cole and kindly made available by them. This algorithm is based on finding spherical regions with a certain predefined mean overdensity $\kappa$. A local density near each particle is needed to provide an initial list of possible halo centres, and is defined as $3(N+1)/(4\pi r_N^3)$, where $r_N$ is the distance to the N'th nearest neighbour. Further details may be found in Lacey & Cole's paper. We follow them in choosing $\kappa = 180$ and $N = 10$. In this algorithm the mass of a halo is simply the number of particles within the bounding sphere.

Many of the statistics we discuss in this paper are based on counts of haloes or of individual particles within randomly placed spheres. When evaluating such statistics for the simulations we use counts of objects within spheres centred on each grid point of a regular $30^3$ cubic mesh. Since the simulations are periodic, there are no difficulties with spheres overlapping the boundary of the simulated region.

### 3.1 Cross-correlation between haloes and mass in Lagrangian space

Our estimate of the average Lagrangian cross-correlation $\bar{\xi}_{\rm hm}^{\rm L}$, given in equation (15), requires no assumptions beyond those of the extended PS formalism. It thus makes a good point to start testing our model. In order to calculate $\bar{\xi}_{\rm hm}^{\rm L}$ in the simulations, we select dark halos at some time $a > 1$. The Lagrangian position of a halo is then taken to be the centre of mass of the initial *unperturbed* positions of its constituent particles. Counts of these halo positions can be computed within spheres of given Lagrangian radius $R_0$. Only spheres with extrapolated linear overdensity less than $\delta_c$ at the time of halo identification are used when calculating the relevant averages since, by hypothesis, spheres with larger $\delta_0$ are part of collapsed dark haloes, and are excluded from our model by the definition of $q$ in equation (9). Figure 1 shows the ratio $\bar{\xi}_{\rm hm}^{\rm L}(R_0)/\bar{\xi}_{\rm m}^{\rm L}(R_0)$ as a function of $\log(R_0/L)$, where $\bar{\xi}_{\rm m}^{\rm L} \propto \Delta^2$ is the average autocorrelation of the extrapolated linear overdensity and is computed directly from the linear power spectrum using equation (2).

The different symbols in figure 1 refer to halos of different mass as detailed in the caption. The corresponding model predictions are shown as the solid curves and are obtained from equation (15) by integrating $n(M_1, z_1)$ and $\mathcal{N}$ over the appropriate range of halo mass $M_1$. The agreement between the model and the simulation results is good for all except the most massive haloes. The problem with the latter almost certainly arises from poor statis-tics since the simulations contain relatively few such haloes. For each power index $n$ and expansion factor $a$, more massive haloes are more strongly correlated with linear overdensity. Haloes with $M < M_*$ are anticorrelated with linear overdensity as expected from equation (13). The agreement found here is not surprising since Lacey & Cole (1994) already showed many aspects of the extended PS formalism to be in excellent agreement with simulations similar to our own.

### 3.2 The bias relation

The second major ingredient of our theoretical formalism is the spherical model which we use to relate the Eulerian quantities $R$ and $\delta$ to the Lagrangian quantities $R_0$ and $\delta_0$. To test this model let us define a bias function $b$ through the relation

$$\delta_h(1|0) = b(R, \delta, M_1, z_1)\delta. \qquad (29)$$

Equation (19) shows that this function depends only on the extended PS formalism and on the spherical collapse model.

In Figure 2, we show $\log(1+\delta_h)$ as a function of $\log(1+\delta)$ for several values of $R/L$ The vertical error bars give the *rms* scatter of the $\delta_h$ values in the corresponding $\delta$ bins. In our simulations the Lagrangian radius of a halo of mass $M$ is about $0.02(M/32)^{1/3}L$. The smallest radius for which results are shown in Figure 2 corresponds to the Lagrangian radius of the least massive haloes considered ($M = 32$). This radius is also smaller than the scales where the mass correlation function becomes nonlinear. It is remarkable that the analytic model reproduces the simulation results over a wide range of $R$ and over almost the entire range of $\delta$ that is probed by the present simulations. In the analytic model, the value of $\delta_h$ drops abruptly to $-1$ when $\delta$ becomes so small that the total mass contained in a sphere is smaller than the mass of the smallest haloes. In the simulations the drop is less abrupt because some of the member particles of a halo can be outside the sphere that contains its centre. For the largest values of $\delta$ statistics are poor in the simulations because very few of our $30^3$ spheres have such values. The scatter in $\delta_h$ among spheres with the same $\delta$ is generally smaller than the mean value of $\delta_h$. Exceptions occur for small spheres and for "almost empty" spheres. Thus it is a reasonable approximation to treat $\delta_h$ as a *function* of $\delta$ as in equation (29). The increase of $\delta_h$ with $\delta$ is faster for haloes with higher $M/M_*$, showing again that haloes with high $M/M_*$ are biased toward regions with large mass overdensities.

Figure 3 shows the same thing as Fig.2, except that haloes are selected at an earlier epoch (when the expansion factor is $a = a_1$) than the one when the $\delta_h$-$\delta$ relation is examined (at $a = a_2 > a_1$). In general, haloes identified at $a_1$ will, by $a_2$, have increased their mass by accretion or lost their identity by merging. However, galaxies which were forming at their centres at $a_1$ may still remain distinct at $a_2$. Thus the relation examined in Figure 3 may be relevant to the biasing of galaxies. In the simulations the position of each halo (or "galaxy") at the later epoch is assumed to be that of the particle which was closest to its centre at $a_1$. The horizontal error bars in Figure 3 represent the scatter in $\delta$ when the count data are rebinned according to the value of $\delta_h$. These bars are also relatively narrow, showing that it may be possible to use "galaxy" overdensities (i.e. $\delta_h$) to predict mass overdensities (i.e. $\delta$). The model predictions in Figure 3 are obtained from equation (19) with $\delta_1$ in $n$ (eq.9) and in $\mathcal{N}$ (eq.12) taking the value $\delta_1 = (a_2/a_1)\delta_c$. Clearly the model also works gratifyingly well in this double epoch context. We conclude that our formalism provides a useful description of the bias function $b(R, \delta, M_1, z_1)$.

### 3.3 Cross-correlation between haloes and mass in Eulerian space

The simplest of the Eulerian correlation statistics for dark haloes is the average cross-correlation between haloes and mass. This quantity depends only on the *mean* number of dark haloes in spheres of given radius $R$ and mass overdensity $\delta$, and it is independent of the scatter in this number. Because of this simplification we begin our comparison of simulation and analytic results for correlations with this statistic. Unfortunately to calculate *any* Eulerian correlation statistic for dark haloes, it is necessary to make some assumption about the probability distribution of the mass overdensity $\delta$. As noted above, this is a significant problem because there are currently no theoretical models for this PDF which remain accurate in the nonlinear regime, $\Delta > 1$. In the following we either take the PDF directly from the simulation itself or use the simple lognormal approximation of equation (22).

Figure 4 shows the ratio of the average cross-correlation between haloes and mass to the average autocorrelation of the mass. The heavy ticks on the horizontal axis show the values of $R$ where $\bar{\xi}_{\rm m} = 1$. Open symbols give results for haloes in various mass ranges as identified by the FOF group finder. Filled dots in Figure 4a give corresponding results for the SO group finder. We see clearly that the two group finders give similar results for the average cross-correlation function $\bar{\xi}_{\rm hm}$. For the case $n = -1.5$, the SO haloes have slightly higher $\bar{\xi}_{\rm hm}$ than the corresponding FOF haloes because the SO group finder breaks clusters into subgroups more efficiently than the standard FOF group finder; this property starts to become significant in a model with substantial large-scale power.

Moving to our analytic predictions, the solid curves in Fig.4 show results from equation (21) using a PDF derived directly from the simulations. These curves match the simulation results in most cases, but not for low $M/M_*$ when $\bar{\xi}_{\rm m} > 1$. Haloes with such low $M/M_*$ are biased toward underdense regions in the Lagrangian space (Fig.1) and our spherical collapse model may be an inadequate description of the nonlinear evolution of underdense regions. The dashed curves in Figure 4a show predictions based on the lognormal PDF of equation (22). When $\bar{\xi}_{\mathrm{m}} \lesssim 1$ these predictions agree with those based on the empirical PDF, but in the nonlinear regime the lognormal predictions are, in general, significantly worse. A better analytic model for the PDF would clearly improve our ability to make predictions in the nonlinear regime. Notice, however, that both predictions remain accurate to values of $\bar{\xi}_{\mathrm{m}}$ exceeding unity. This often corresponds to $\bar{\xi}_{\mathrm{hm}}$ values much above unity.

In Figure 4b we show " two epoch" cross-correlations between haloes and mass. As in Figure 3 the haloes are identified at $a_1$ but the correlations are calculated at the later epoch $a_2 > a_1$. In this case the agreement between the analytic model (using the empirical PDF) and the simulation results is excellent even on scales where the mass autocorrelation is substantially nonlinear. This reinforces the idea that careful use of our analytic techniques should allow an accurate prediction of galaxy correlations for any specific model of the relation between galaxies and the haloes in which they form.

### 3.4 Autocorrelations of haloes in Eulerian space

Estimating average autocorrelations for dark haloes requires knowledge of the scatter in the number of haloes in spheres of given radius and mass overdensity in addition to knowledge of the mean number. As discussed in §2.4, we have not yet found an adequate way to calculate this within our formalism. We now demonstrate explicitly that improper treatment of halo exclusion effects can cause serious errors in the estimate of average autocorrelations, but that a simpler linear bias model for the standard autocorrelation function actually works quite well.

In Figure 5 we use individual symbols to show the ratio $\sigma_{\mathrm{hh}}^2(R, M_1, z_1)/\bar{\xi}_{\mathrm{m}}(R)$ as a function of $R$ for haloes identified using the FOF group finder. Dashed curves in the left-hand panels show corresponding results for SO haloes. On small scales the SO values are systematically higher than the FOF values, particularly for massive haloes. As before, this is because the SO group finder breaks clusters into subgroups more efficiently than the FOF algorithm. Solid curves in this figure show the predictions of equation (23). The discrepancies are substantial on small scales, and indeed on all scales for $n = 0$. Count variances in the simulations increase less rapidly to small scales than either $\bar{\xi}_{\mathrm{m}}$ or $\bar{\xi}_{\mathrm{hh}}$ as given by equation (23). This discrepancy appears due to halo exclusion effects as discussed in §2.4 (see equation 25). This is consistent with the fact that the predictions are better for large $R$, for small $M$, and for haloes identified at an earlier epoch than the one for which the average autocorrelation is calculated (see the right panels of Fig.5). For $n = 0$ clustering on large scales is weak, and the effect of halo exclusion propagates to large scales in $\sigma_{\mathrm{hh}}^2(R)$.

In order to avoid these exclusion problems we now consider the standard autocorrelation function of dark haloes ($\xi_{\mathrm{hh}}(R)$ defined by equation 26) rather than the average autocorrelation function of Figure 5. In Figure 6 we use squares to show this autocorrelation function for haloes identified using the FOF algorithm. The error bars represent bootstrap errors, as discussed in Mo, Jing & Börner (1992). Dashed curves in Fig.6a show the corresponding results for SO haloes; for reasons anticipated in the discussion of §2.4, $\xi_{\rm hh}(R)$ does not depend significantly on the group finder for $R$ larger than the linear sizes of haloes. (The linear sizes are about $0.02L$ and $0.03L$ for haloes with $M = 32$ and $128$, respectively.) The solid curves in Figure 6 show the predictions of equation (28), with $\xi_{\rm m}(R)$ estimated directly from the simulations. In practice, $\xi_{\rm m}$ can equally well be obtained from the initial density power spectrum by using the fitting formula of Jain et al. (1995); this procedure allows a purely analytic estimation of halo autocorrelations. It is clear from Fig.6 that this simple model fits the simulation results quite well over a wide range of scales. The thick ticks on the horizontal axes mark the values of $R$ where $\bar{\xi}_{\rm m}(R) = 1$. Our model can work well even for $\bar{\xi}_{\rm m} > 1$; this is particularly the case when autocorrelations are estimated for haloes identified at some earlier epoch. On the other hand, the model can break down on scales smaller than the linear sizes of haloes because of the exclusion effects discussed above.

## 4 DISCUSSION

The tests of the last section show that the spatial distribution of dark haloes and its relation to the underlying mass distribution can be described quite accurately by our simple analytic model. As a result this model should provide some indication of how the observed galaxy distribution may related to that of the mass. In this section, we will briefly discuss several possible applications of these results. Details of these applications will be described elsewhere.

As a first application consider how the galaxy formation process may bias the galaxy distribution. The results presented in Figures 2 and 3 show clearly that the bias of dark haloes depends not only on their mass but also on the epoch when they are identified. Indeed, for haloes of a given mass, the bias increases strongly with the redshift of identification. Thus the objects which formed at the centres of early, relatively low-mass haloes can be more strongly clustered today than current haloes of larger mass. This result is interesting. In a hierarchical clustering scenario, such as the CDM model or any of its currently popular variants, high mass haloes form through the merger of smaller systems. If these early haloes produced galaxies which survive to the present day, these galaxies could be just as strongly clustered as the more massive galaxies which form later. A strong mass (or luminosity) dependence of galaxy correlations, in the sense that brighter galaxies are more strongly clustered, is not a necessary consequence of the hierarchical model. Our model predicts that objects which form at redshift $z$ in haloes with mass $M = M_*(z)$ will be unbiased relative to the mass at all later redshifts (equation 20). A strong "natural" bias in the galaxy distribution of the kind advocated by White et al. (1987) can only be achieved if most galaxies formed at the centres of haloes with $M > M_*$. At any given time massive dark haloes may, of course, contain more than one galaxy, and some low mass haloes may contain no galaxies at all. The observed galaxies do not, presumably, correspond uniquely to the centres of the haloes present at any single epoch. As a result, it is not straightforward to apply our results directly to galaxies. However, if more detailed modelling allows a prediction of the number of galaxies in a halo as a function of its mass and of galaxy properties (e.g. Kauffmann, White & Guiderdoni 1993; Kauffmann, Guiderdoni & White 1994; Cole et al. 1994) our results can readily be extended to study galaxy clustering as a function of luminosity, morphological type, colour, or any other property of interest.

A second application is to the study of how the bias function $b(R, \delta, M_1, z_1)$ depends on $\delta$ and $R$. In particular, our model should enable us to see the extent to which the assumption of linear bias (i.e. $b = \text{const.}$) is valid. In Figure 7 we show model predictions for the $\delta_\text{h}$-$\delta$ relation for spheres with $R = 0.05L$. We see clearly that the relation can deviate from the simple linear one, $\delta_\text{h} \propto \delta$. The deviation is larger for haloes identified at later epochs, for more massive haloes, and for smaller values of $R$. To see this dependence more clearly, let us expand $\delta_\text{h}$ to second order in $\delta$ and to first order in $\eta \equiv (\Delta_0/\Delta_1)^2$. From equation (18), we have $\delta_0 = \delta + c\delta^2$ with $c = -0.805$. Expanding equation (19) to the necessary order gives

$$\delta_\text{h} = B_0 + B_1 \delta + \frac{1}{2} B_2 \delta^2, \qquad (30)$$

where $B_0 = (\eta/2)(3 - \nu_1^2)$ is a zero-point offset, $B_1 = 1 + (\nu_1^2 - 1)/\delta_1$ is the linear bias of equation (20), and $B_2 = (\nu_1/\delta_1)^2(\nu_1^2 - 3) + 2(\nu_1^2 - 1)(1 + c)/\delta_1$. The second order coefficient $B_2$ can be either positive or negative depending on the value of $\nu_1^2$. In particular, for haloes of mass $M_1 < M_*(z_1)$ we have $\nu_1 < 1$ and $\delta_\text{h}$ always falls below the linear bias relation for large $\delta^2$. This negative second derivative is quite evident in the curves of Figure 7, as is the zero-point offset which is largest for massive haloes.

Relations such as equation (30) may also have important implications for the skewness and other high-order moments of the halo count distribution. If we could neglect the scatter in halo counts for spheres of given $R$ and $\delta$, then the formulae of Fry & Gaztanaga (1993) would allow the skewness of the halo counts to be written as $S_h = (S_m + 3B_2/B_1)/B_1$, where $S_m$ is the skewness of the PDF of $\delta$. For $M_*$ haloes identified at redshift $z_1$ this gives $S_h = S_m - 6/\delta_1^2$, which is substantially smaller than $S_m$ unless $z_1$ is high. While additional contributions to the skewness of halo counts come from scatter terms analogous to the $\mu(R, \delta)$ term in equation (25), this reduction of $S_h$ caused by the nonlinearity of the mean bias relation may explain why late-type galaxies have a lower value of skewness than early-type galaxies (Jing, Mo, Börner 1991). This line of argument is obviously worth further investigation.

The results of §3.4 show that to a good approximation the halo autocorrelation function is directly proportional to the mass autocorrelation function with a constant bias which can be calculated simply using our model [equations (20) and (28)]. This validates the usual linear bias assumption well into the nonlinear regime, and shows directly how the bias is related to the "peak height" of haloes through the parameter $\nu_1$. From equation (6) we see that this parameter also determines the fraction of the cosmic mass in haloes with masses exceeding $M_1$. If the abundance of a certain type of haloes is known as a function of their mass, then this fraction can be estimated, and so $\nu_1$ calculated, for any assumed value of the cosmic density parameter $\Omega_0$. This in turn allows the mass correlation function to be estimated directly from the observed halo correlation function. Thus the conventional normalization amplitude for mass fluctuations $\sigma_8$ is obtained, again as a function of the assumed $\Omega_0$. One can then try different values of $\Omega_0$ and require that the $\sigma_8$ obtained from this argument is consistent with other determinations [e.g. from large-scale flows (Dekel 1994), from cluster abundances (White, Frenk & Efstathiou 1993) and from weak gravitational lensing (Villumsen 1995)]. In this way one may hope to measure $\Omega_0$.

**Acknowledgements:**

We thank S. Cole and C. Lacey for providing us with their code for the SO group finder. This research was supported in part by the National Science Foundation under Grant No. PHY89-04035 to the Institute for Theoretical Physics at Santa Barbara.

# REFERENCES

Bardeen J., Bond J.R., Kaiser N., Szalay A.S., 1986, ApJ, 304, 15
Bernardeau F., 1994, A&A, 291, 697
Bond J.R., Cole S., Efstathiou G., Kaiser N., 1991, ApJ, 379, 440
Bond J.R., Meyers S., 1995a,b, CITA Preprints
Bower R.J., 1991, MNRAS, 248, 332
Cole S., Aragon A., Frenk C.S., Navarro J.F., Zepf S., 1994, MNRAS, 271, 781
Cole S., Kaiser N., 1989, MNRAS, 237, 1127
Coles P., Jones B., 1991, MNRAS, 248, 1
Colombi S., 1994, ApJ, 435, 536
Davis M., Efstathiou G., Frenk C., White S.D.M., 1985, ApJ, 292, 371
Dekel A., 1994, ARA&A, 32, 371
Efstathiou G., Frenk C.S., White S.D.M., Davis M., 1988, MNRAS, 235, 715
Frenk C.S., 1991, Physica Scripta, T36, 70
Frenk C.S., White S.D.M., Davis M., Efstathiou G., 1988, ApJ, 327, 507
Fry J.N., Gaztanaga E., 1993, ApJ, 413, 447
Gelb J.M., Bertschinger E., 1994a, ApJ, 436, 467
Gelb J.M., Bertschinger E., 1994b, ApJ, 436, 491
Jain B., Mo H.J., White S.D.M., 1995, MNRAS, 276, L25
Jing Y.P., Mo H.J., Börner G., 1991, A&A, 252, 449
Kaiser N., 1984, ApJL, 284, L9
Katz N., Quinn T., Gelb J.M., 1993, MNRAS, 265, 689
Kauffmann G., Guiderdoni B., White S.D.M., 1994, MNRAS, 267, 981
Kauffmann G., White S.D.M., 1993, MNRAS, 261, 921
Kauffmann G., White S.D.M., Guiderdoni B., 1993, MNRAS, 264, 201
Kolb W., Turner M.S., 1990, The Early Universe, Addison-Wesley, Reading
Lacey C., Cole S., 1993, MNRAS, 262, 627
Lacey C., Cole S., 1994, MNRAS, 271, 676
Mo H.J., Jing Y.P., Börner G., 1992, ApJ, 392, 452
Peebles P.J.E., 1980, The Large-Scale Structure of the Universe, Princeton University Press, Princeton
Press W.H., Schechter P., 1974, ApJ, 187, 425 (PS)
Villumsen J., 1995, MPA Preprint
White S.D.M., 1995, in Schaeffer R., ed., 1993 Les Houches Lectures, in press
White S.D.M., Davis M., Efstathiou G., Frenk C.S., 1987, Nat, 330, 451
White S.D.M., Frenk C.S., 1991, ApJ, 379, 52
White S.D.M., Frenk C.S., Efstathiou G., 1993, MNRAS, 262, 1023
White S.D.M., Rees M.J., 1978, MNRAS, 183, 341
20

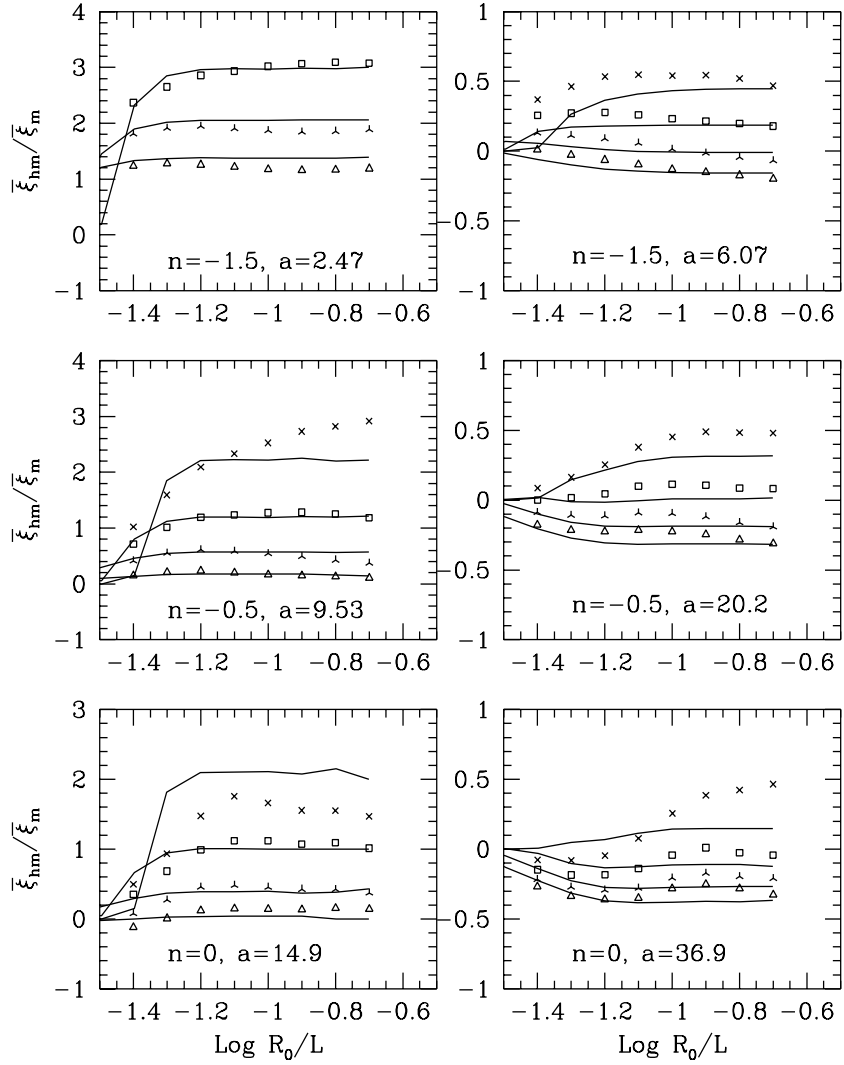

**Figure 1:** The average cross-correlation function between haloes and mass in Lagrangian space, normalized by the linear average mass correlation function. Results from simulations are shown for haloes with mass $M \geq 32$ (triangles), 64 (three-pointed stars), 128 (squares), and 256 (crosses). The solid curves are the model predictions. The statistics for the $M > 256$ samples in the simulations are poor, because they contain only small number of haloes. The power index $n$ and the expansion factor $a$ are shown in each panel.

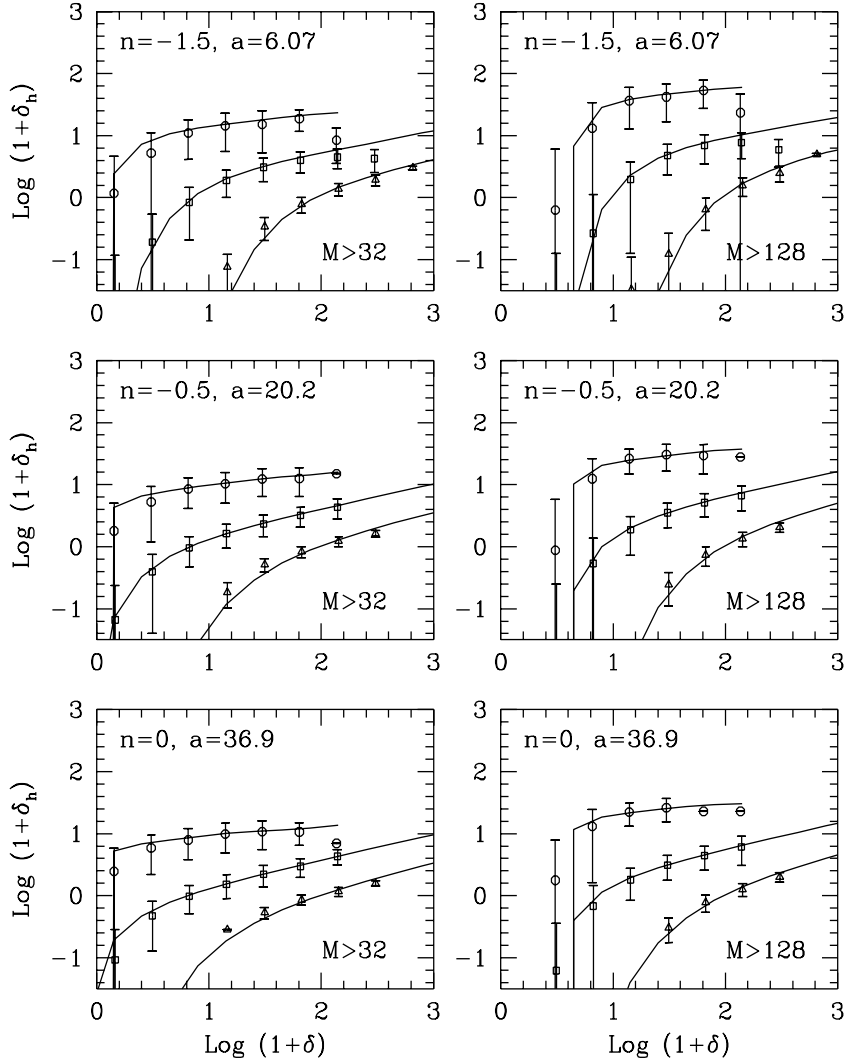

**Figure 2a:** The bias relation, i.e. the overdensity of haloes $\delta_h$ versus the overdensity of mass $\delta$ in spherical windows with radius $R$. Circles, squares and triangles show the simulation results for $R/L = 0.02$, $0.05$ and $0.13$, respectively. To avoid crowding, the results for $R/L = 0.05$ and $0.13$ are shifted by 1 and 2 decades along the horizontal axis. Vertical error bars show the $1\sigma$ scatter of the $(1+\delta_h)$ values in the corresponding $\delta$ bins. Model predictions are shown by the solid curves. Results are shown for haloes with $M > 32$ (left panels) and $M > 128$ (right panels). Here haloes are selected at the same epoch $a$ as when the correlation function is calculated. The power index $n$ and the expansion factor $a$ are shown in each panel.

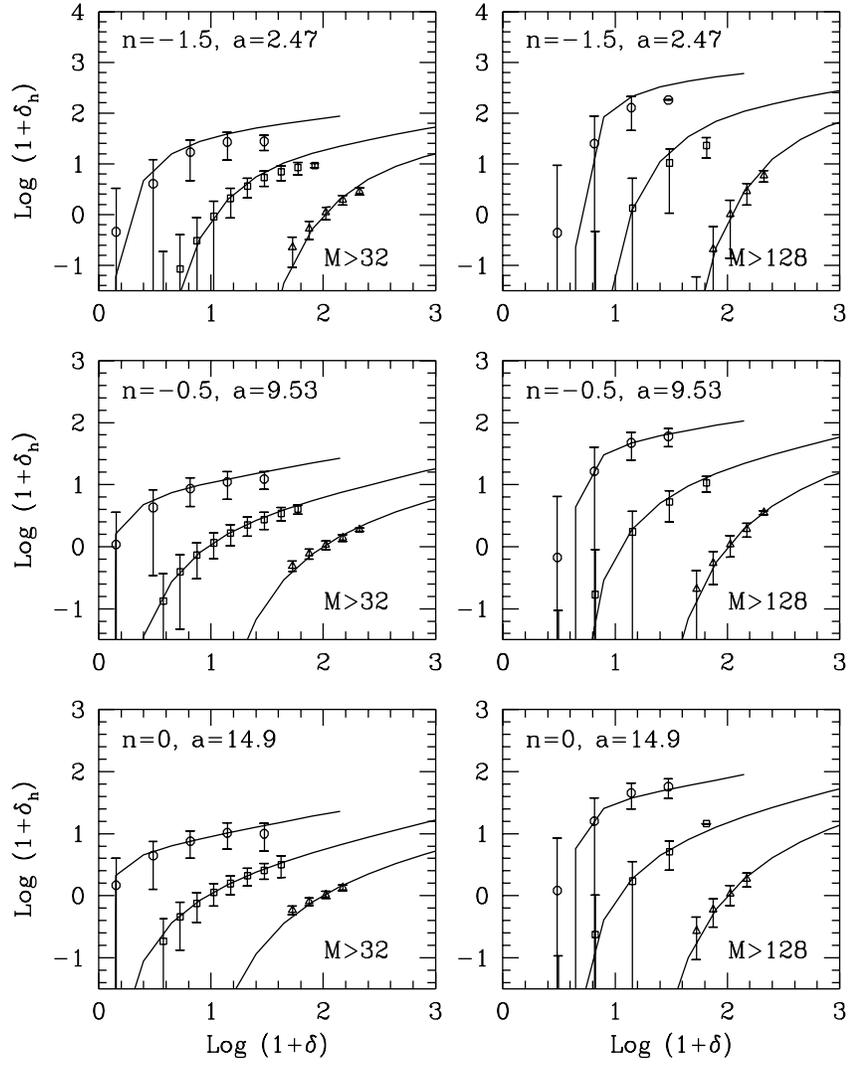

**Figure 2b:** The same as Figure 2a for a different set of expansion factors.

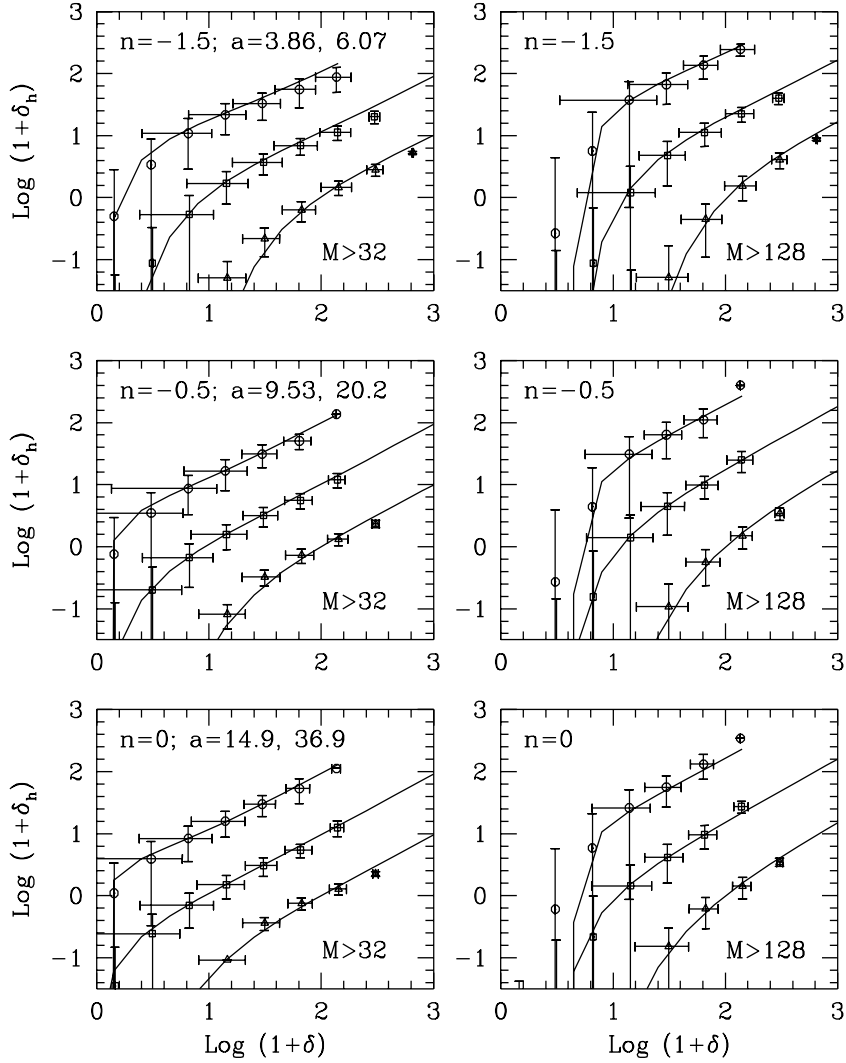

**Figure 3:** The same as Figure 2, but here haloes are selected at an earlier epoch (when the expansion factor $a$ has the lower value indicated in each panel) than when the $\delta_h$-$\delta$ relation is examined (at the epoch with the higher $a$). The horizontal error bars represent the scatter of $\delta$ among spheres within the corresponding $\delta_h$ bins.

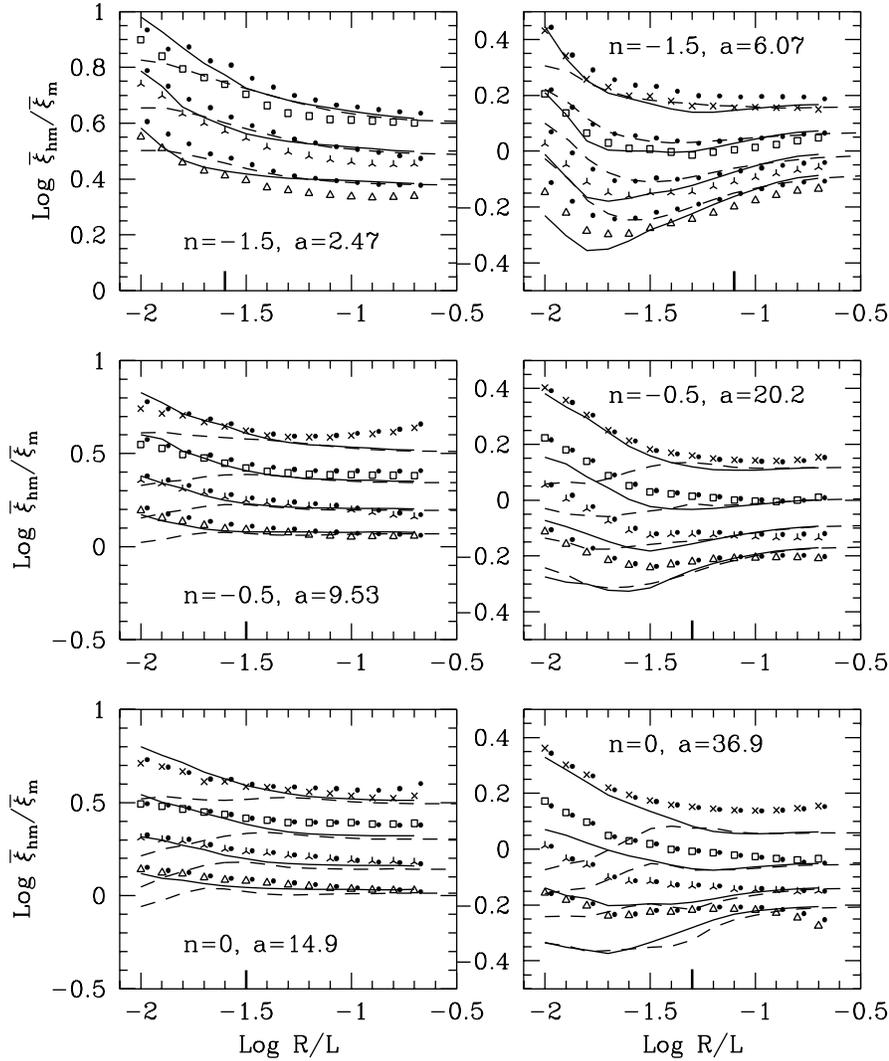

**Figure 4a:** The average cross-correlation function between haloes and mass, normalized by the autocorrelation function of the mass. Triangles, three-pointed stars, squares and crosses show the results of simulations for haloes, identified by the standard FOF group finder, with mass $M \geq 32$, 64, 128, and 256, respectively. The solid circles show the corresponding results for haloes identified by the SO group finder. The solid curves show the model predictions, with the PDF derived directly from the simulations. The dashed curves show the results obtained from the lognormal PDF. The thick ticks on the horizontal axis show the values of $R$ where the average mass correlation function $\bar{\xi}_m = 1$. Here haloes are selected at the same epoch $a$ as when the correlation function is calculated. The power index $n$ and the expansion factor $a$ are shown in each panel.

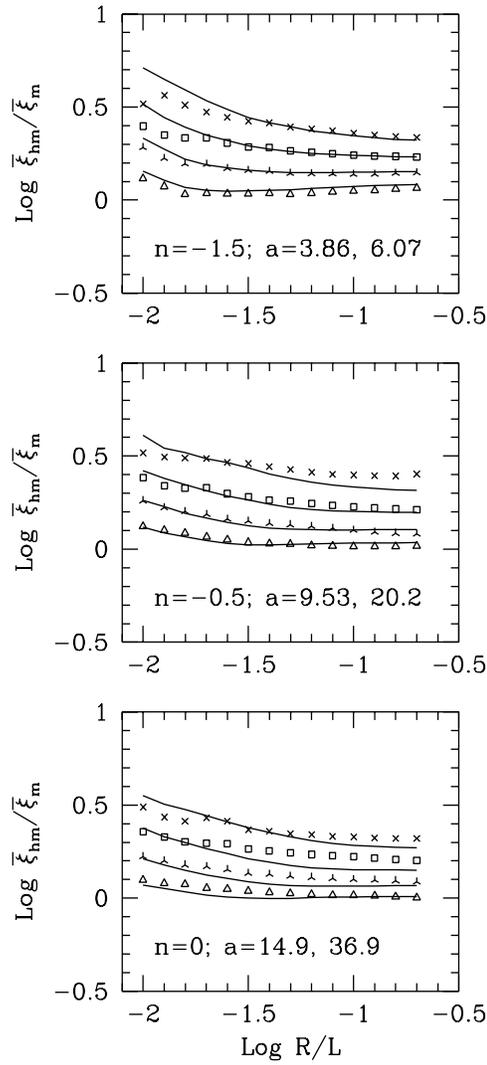

**Figure 4b:** The same as 4a, but here haloes are selected at an earlier epoch (when the expansion factor $a$ has the lower value indicated in each panel) than when the correlation function is calculated (at the epoch with the higher $a$).

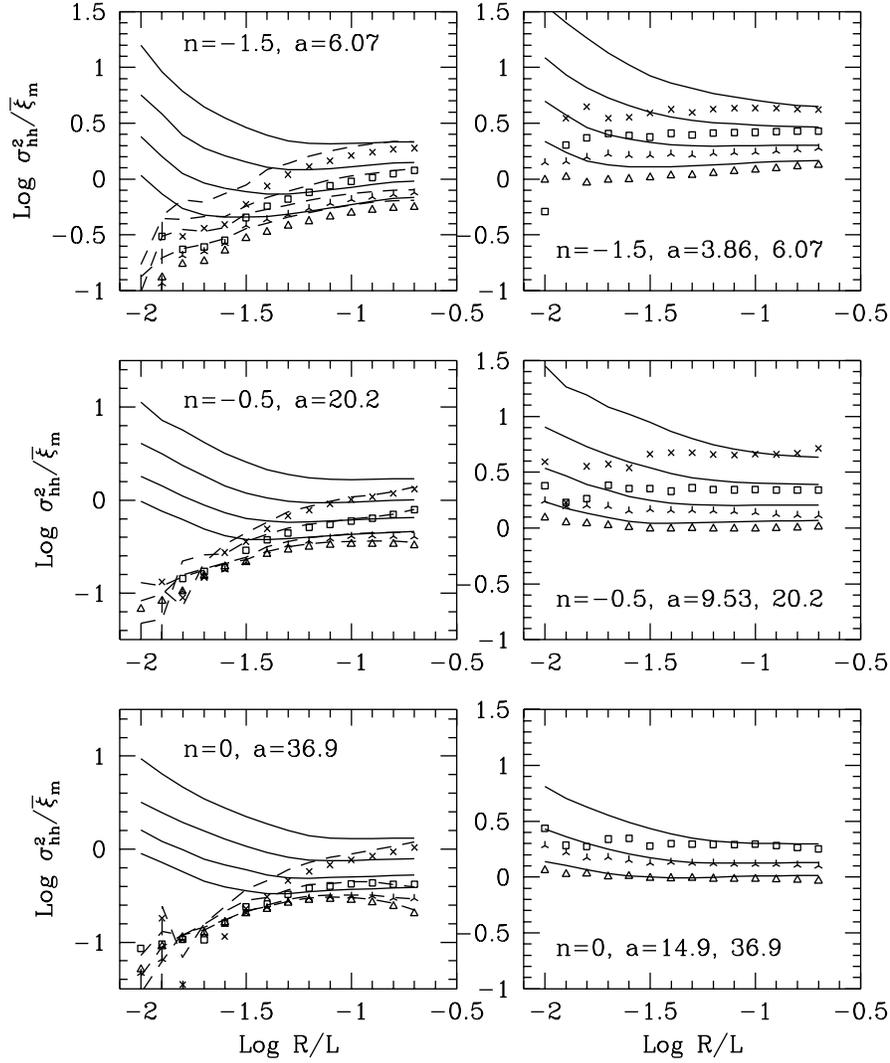

**Figure 5:** The variances of halo counts in the simulations. Symbols show the results for FOF haloes with $M > 32$ (triangles), 64 (three-pointed stars), 128 (squares) and 256 (crosses). Dashed curves (in the left panels) show the results for the corresponding SO haloes. The solid curves show the predictions of equation (23). The results shown in the right panels are for haloes selected at an earlier epoch (when the expansion factor $a$ has the lower value indicated in each panel) than when the correlation function is calculated (at the epoch with the higher $a$).

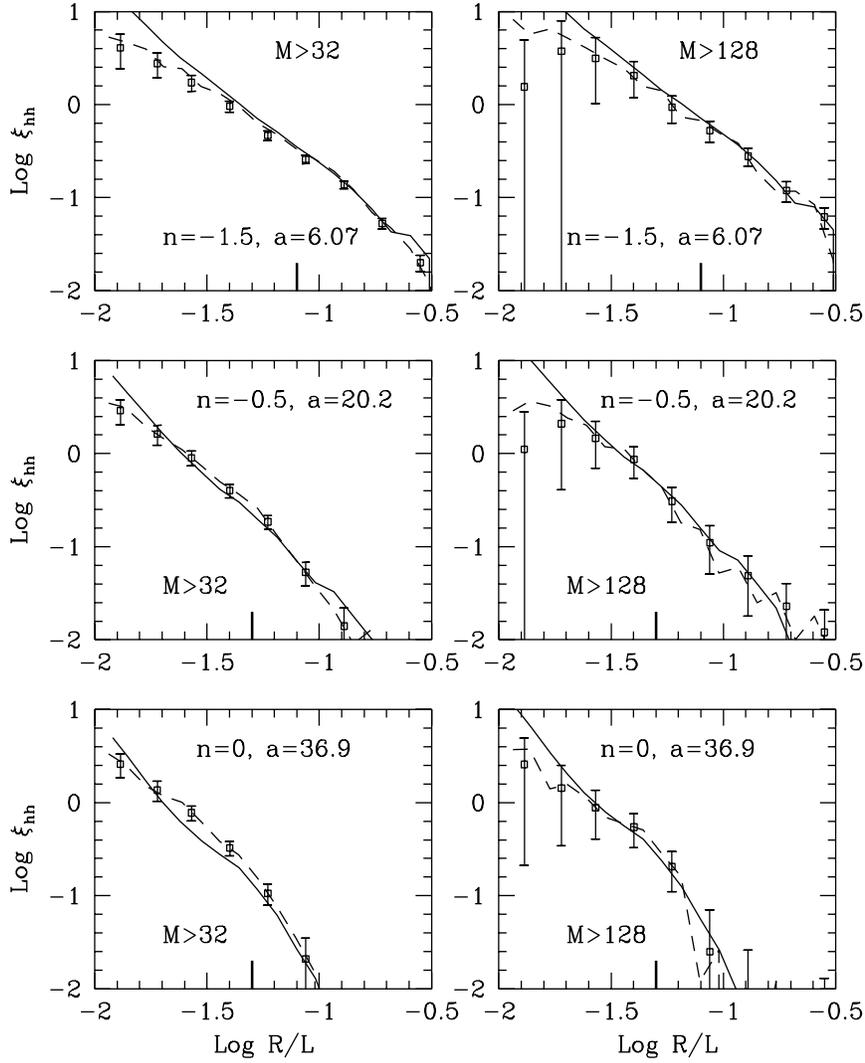

**Figure 6a:** The standard two-point correlation functions of haloes in the simulations. Squares are for FOF haloes, while dashed curves for SO haloes. The error bars represent bootstrap errors. The solid curves show the predictions of equation (28). The thick ticks on the horizontal axes show the values of $R$ where $\bar{\xi}_{\mathrm{m}} = 1$. Results are shown for haloes with mass $M > 32$ (left panels) and $M > 128$ (right panels). Here haloes are selected at the same epoch $a$ as when the correlation function is calculated.

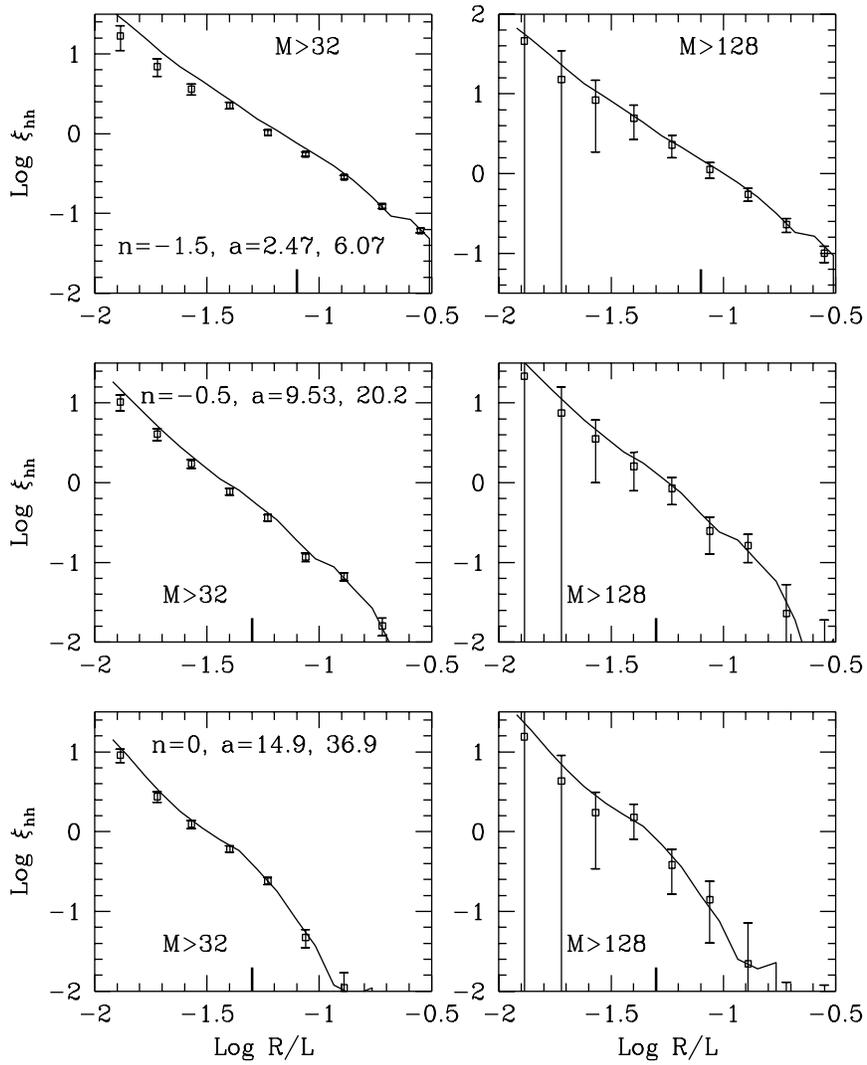

**Figure 6b:** The same as 6a, but here haloes are selected at an earlier epoch (when the expansion factor $a$ has the lower value indicated in each panel) than when the correlation function is calculated (at the epoch with the higher $a$).

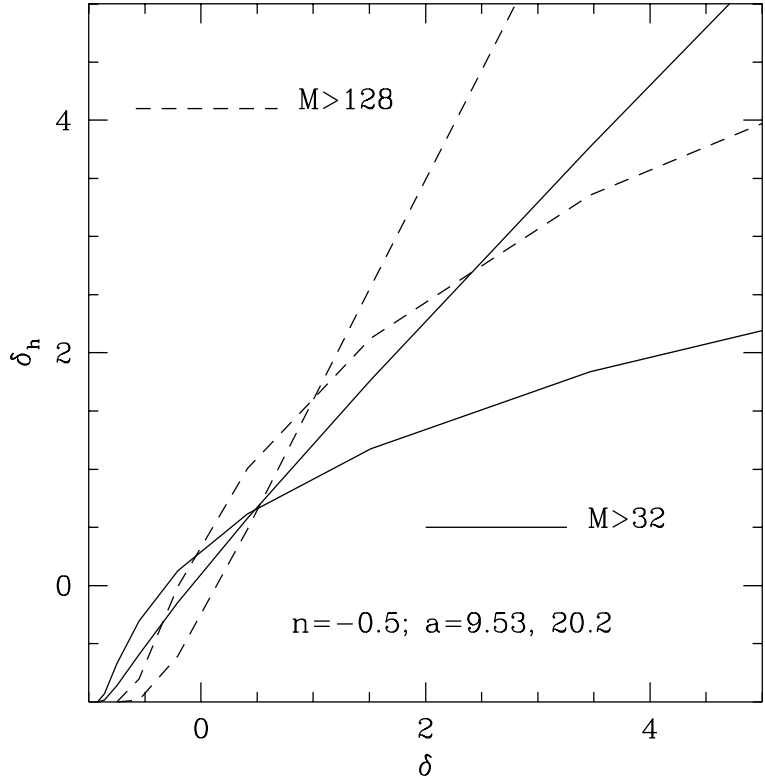

**Figure 7:** The model predictions of the bias relation for spheres with $R = 0.05L$, plotted in linear coordinates to show the deviation from the linear relation $\delta_h \propto \delta$. Results are shown for haloes with mass $M > 32$ and $M > 128$. For a given mass, the steeper curve shows the results for haloes identified at $a = 9.35$ and analyzed at $a = 20.2$, while the shallower one shows the results for haloes both identified and analysed at $a = 20.2$.